\documentclass{article}
\usepackage{graphicx,amsmath,subfigure}
\usepackage[english]{babel}

\title{
\includegraphics[width=0.35\textwidth]{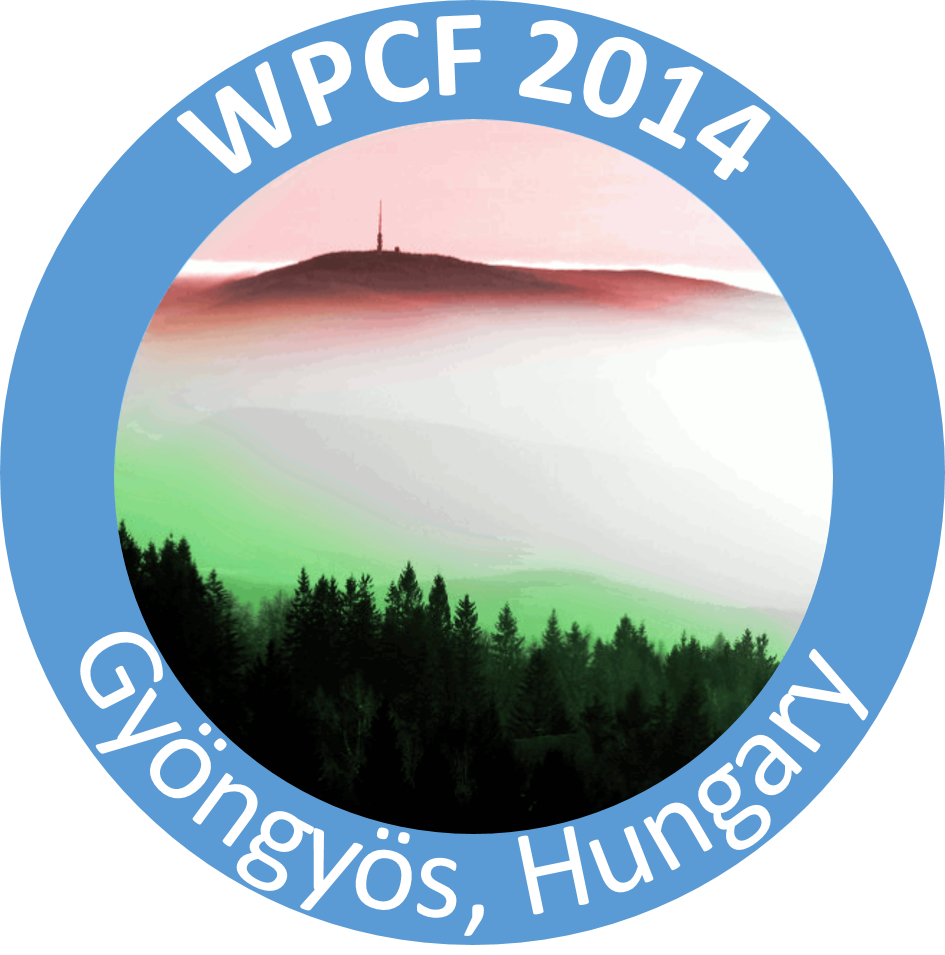}\\[1cm]
Higher order anisotropies in hydrodynamics}
\author{{M. Csan\'ad, A. Szab\'o, S. L\"ok\"os, A. Bagoly,}\\[1ex]
E\"otv\"os University, P\'azm\'any P. s. 1/a, 1117 Budapest, Hungary\\
}

\begin{document}

\fontfamily{lmss}\selectfont
\maketitle

\begin{abstract}
In the last years it has been revealed that if measuring relative to higher order event planes $\Psi_n$, 
higher order flow coefficients $v_n$ for $n>2$ can be measured. It also turned out that Bose-Einstein (HBT)
correlation radii also show 3rd order oscillations if measured versus the third order event plane $\Psi_3$.
In this paper we investigate how these observables can be described via analytic hydro solutions and hydro
parameterizations. We also investigate the time evolution of asymmetry coefficients and the mixing of
velocity field asymmetries and density asymmetries.
\end{abstract}

\section{Introduction}
In relativistic heavy ion collisions, an expanding and cooling medium is created, usually referred to as 
the strongly interacting quark gluon plasma. Hydrodynamics provides a tool to investigate the time evolution
of this medium, and exact analytic models are particularly useful in this regards. Usually spherical, axial
or ellipsoidal symmetry is assumed in these solutions, as these are simple to handle and represent geometries
that yield realistic results for many soft observables. However, event-by-event fluctuating nuclear
distributions yield event-by-event fluctuating initial conditions, and thus higher order azimuthal asymmetries
arise. In order to describe these one has to utilize higher order asymmetries in hydro as well. This was successfully
done in numerical calculations, see for example~\cite{Chatterjee:2014nta,Yan:2015jma}. 

In this paper we discuss the first exact analytic solutions~\cite{Csanad:2014dpa} of relativistic hydrodynamics that
assume higher order asymmetries and thus give realistic higher order flow coefficients. We also discuss possible
extensions of this approach, by analyzing the time evolution of the asymmetries in a numerical approach, and by
investigating the effect of spatial versus momentum space anisotropies.

\section{Multipole solutions and higher order anisotropies}
The first 1+3D relativistic solution with ellipsoidal geometry was discovered in Ref.~\cite{Csorgo:2003ry}.
In this solution the thermodynamic quantities at a given proper-time $\tau$ are constant on the surfaces of an expanding ellipsoid,
defined by the $s$ scale variable
\begin{align}
s=\frac{r_x^2}{X^2}+\frac{r_y^2}{Y^2}+\frac{r_z^2}{Z^2},\label{e:scale2}
\end{align}
where $r_x$,$r_y$,$r_z$ are the spatial coordinates, while $X$, $Y$, $Z$ are the time-dependent axes of the ellipsoid.
The velocity profile as a function of space-time coordinates  $x^{\mu}$ is given in form of a 3D Hubble flow, i.e.
$u^{\mu}=x^{\mu}/\tau$. With these, $u^\mu\partial_\mu s = 0$ holds (if the expansion of the axes is linear in
time).  In Ref.~\cite{Csorgo:2003ry} it was already indicated, that more complicated
scale variables can also be written up (with $u^\mu\partial_\mu s = 0$ still holding). 
In Ref.~\cite{Csanad:2014dpa} we showed that this solution can indeed be extended to 
multipole symmetries with a generalized scale variable
\begin{align}
s=\frac{r^N}{R^N}\left(1+\epsilon\cos(N\phi)\right)\label{e:sN}
\end{align}

\begin{figure}
\centering
\includegraphics[width=120mm]{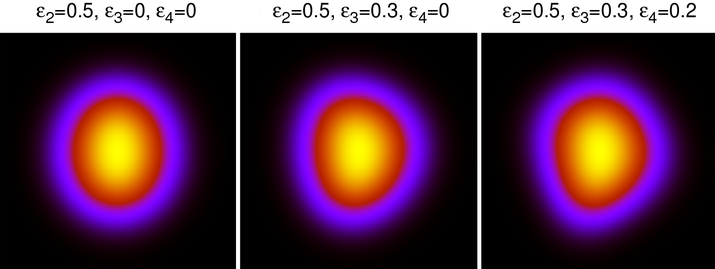}
\caption{Heat map of $s$ values in the transverse plane, with multiple superimposed symmetries. The more $\epsilon_N$
components are included, the more asymmetric the shape gets.}
\label{f:multis}
\end{figure}

With the $s$ given in Eq.~(\ref{e:sN}), the new solutions can be given in cylindrical coordinates $(r,\phi,z)$ as:
\begin{align}
s&=\sum\limits_N \frac{r^N}{R^N}\left(1+\epsilon_N\cos(N(\phi-\psi_N))\right)+\frac{z^N}{R^N}\label{e:sNm}\\
n&=n_f\left( \frac{\tau_f}{\tau} \right)^3 \nu(s)\\
T&=T_f\left(\frac{\tau_f}{\tau} \right)^{3/\kappa}\frac{1}{\nu(s)}\\
p&=p_f\left( \frac{\tau_f}{\tau} \right)^{3+3/\kappa}
\end{align}
and $u^\mu$ still representing a Hubble-flow, as in the original paper of Ref.~\cite{Csorgo:2003ry}.
In the formula for $s$, $\psi_N$ being the $N$th order reaction planes (which cancel from the observables). This way we get
new solutions with almost arbitrary shaped initial distributions, see Fig.~\ref{f:multis}. It is important to 
note here that however, the initial state fluctuation in the observed collision is present through the
orientation of the $N$th order reaction planes and the strength of higher order asymmetries, the
event plane orientation itself does not affect the measured quantities. Thus if every $v_N$ is
measured relatively to the $N$th order reaction plane, then the (event-through-event) averaged
value of $v_N$ will correspond to an average $n$-pole anisotropy $\epsilon_N$.

\begin{figure}
\centering
\includegraphics[width=1\linewidth]{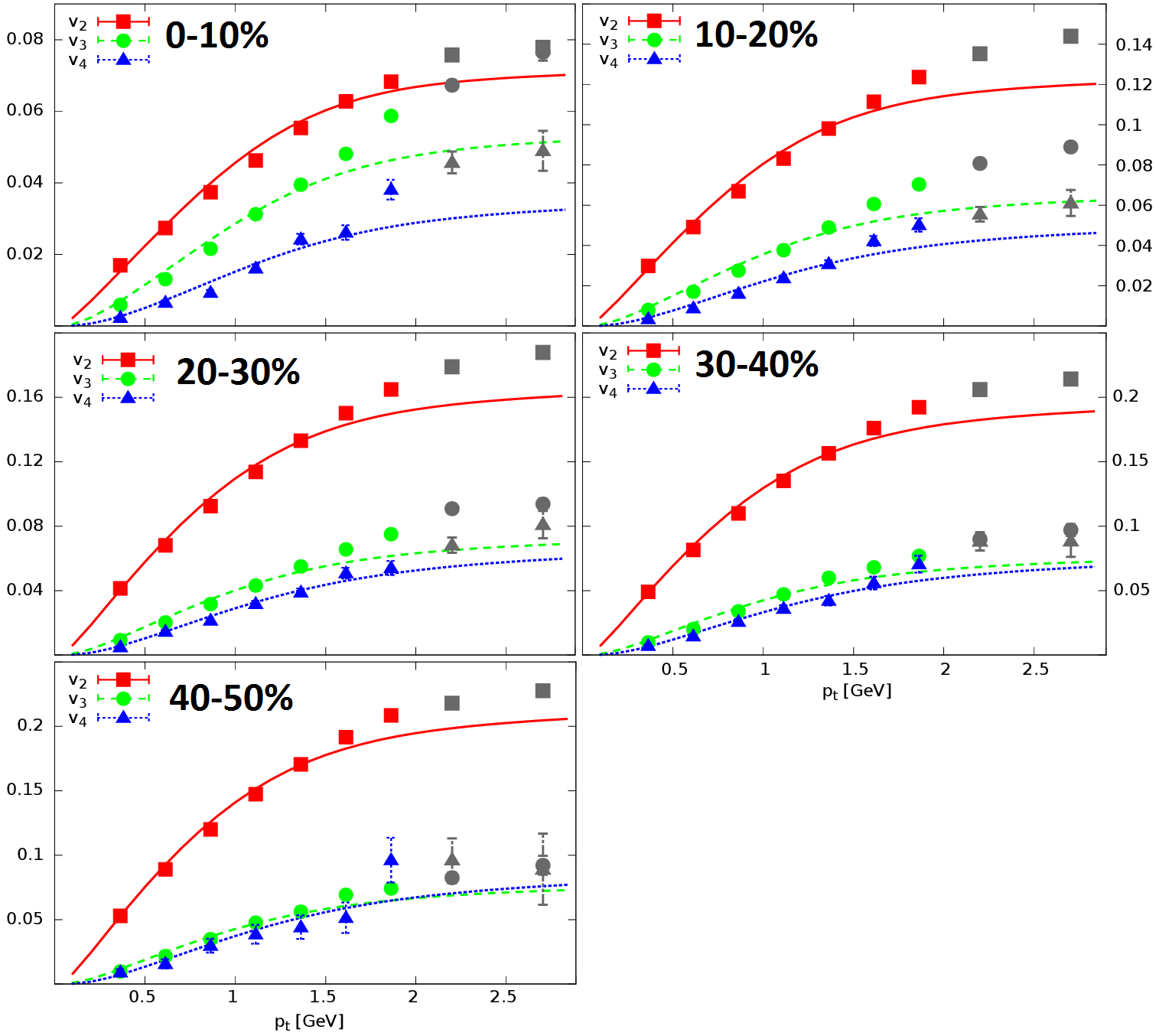}
\caption{Fits to PHENIX 200 GeV Au+Au data~\cite{Adare:2011tg} in 5 centrality bins. Fit parameters are summarized
in Ref.~\cite{Csanad:2014dpa}}
\label{fig:fits}
\end{figure}

We also calculated hadronic observables from the above solution (see details of the freeze-out scenario in Ref~\cite{Csanad:2014dpa} or Ref.~\cite{Csanad:2009wc}. A comparison to PHENIX data on higher order
harmonics measured in 200 GeV Au+Au collisions~\cite{Adare:2011tg} is shown in Fig.~\ref{fig:fits}.
Fit parameters of the model are $\epsilon_N$ (for $N=2,3,4$), $u_t$ and $b$
($T_0$ and $\tau_0$ was fixed to values given from spectra and HBT comparisons of a similar model, as described
in Refs.~\cite{Csanad:2014dpa,Csanad:2009wc}).

\section{Time evolution of the anisotropies}
In the above described solution, the anisotropies don't change over time - due to the lack of pressure gradients and
the Hubble-flow. In a numerical framework, we investigated how the introduction of pressure gradients, various
speeds of sound and viscosity coefficients influence the time
evolution of the asymmetries, when starting from an initial condition that is very similar to one described by know
analytic solutions -- except in pressure, where we used a pressure profile similar to the density profile given in usual
Hubble-expansion models~\cite{Csorgo:2003ry,Csanad:2014dpa}. We used a multi-stage predictor-corrector
method outlined in Ref.~\cite{Toro:2006aa}.	 This is a finite volume scheme, where the initial flux is a
weighted average of the Lax-Friedrichs and Lax-Wendroff fluxes, called GFORCE. This flux is used to
make a new prediction on the grid points, which is used to get a better flux approximation. This procedure is
repeated for a number of times, as described e.g. in Ref.~\cite{Toro:2006aa}. This multi-stage flux
gives results that are comparable to those of the Godunov method. We tested our method with known analytic solutions
given in Refs.~\cite{Csorgo:2001xm,Csorgo:2003ry,Csanad:2014dpa}.
We analyzed both non-relativistic and relativistic hydrodynamics, and arrived at similar conclusions.

\begin{figure}
\begin{center}
    	\includegraphics[width=1\textwidth]{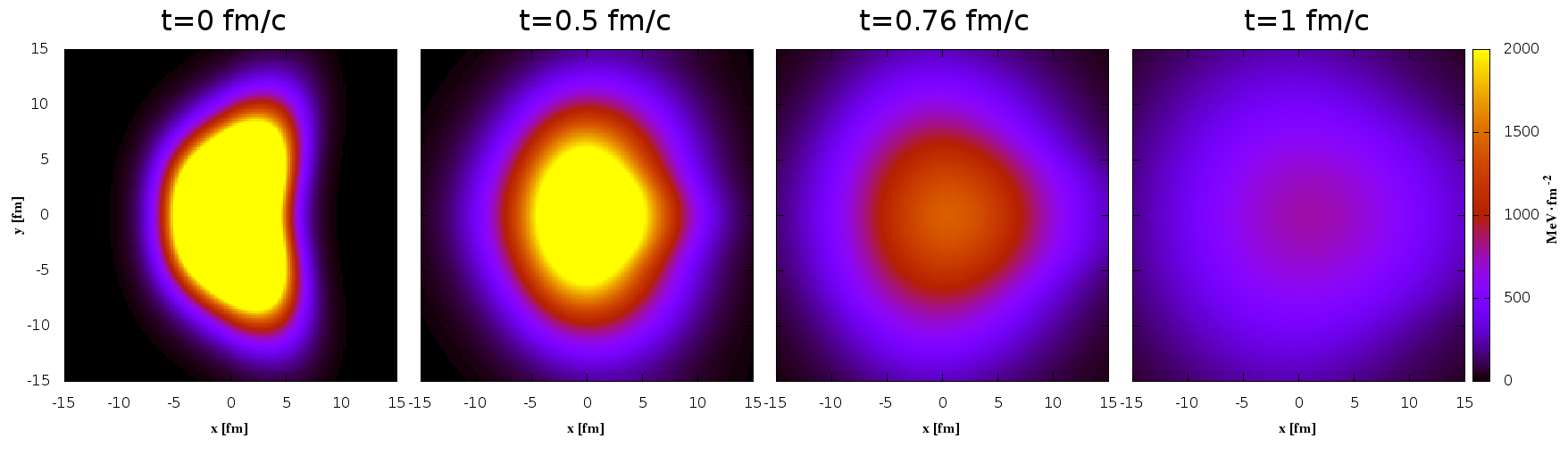}
    	\includegraphics[width=1\textwidth]{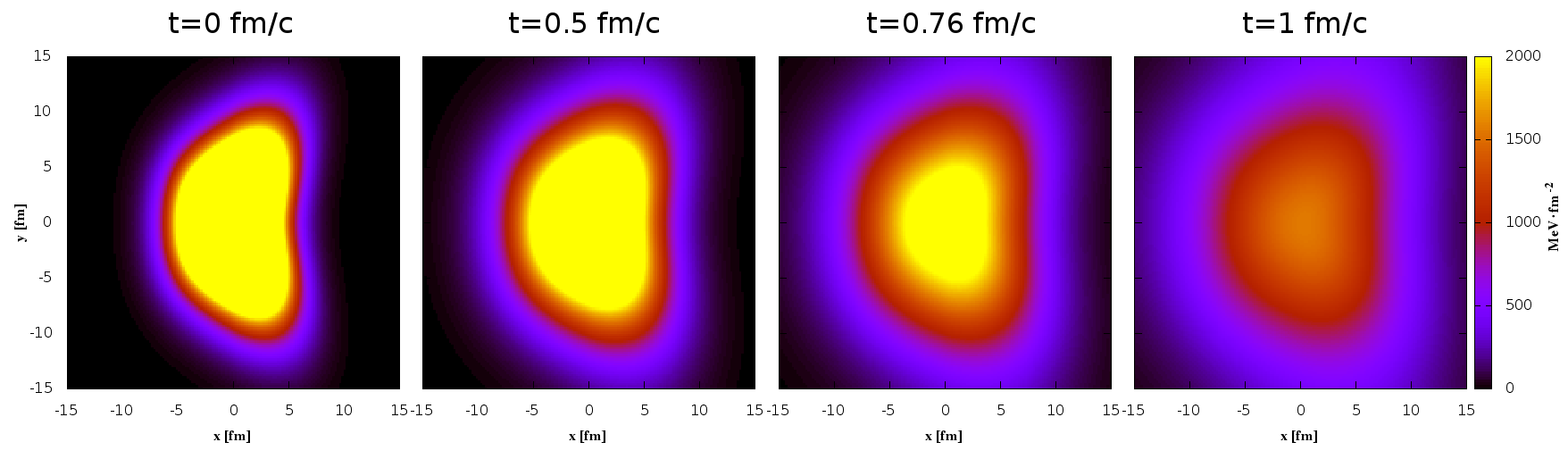}
	\caption{Time evolution of energy density. Top row shows viscosity free case, the bottom row with $\mu={\rm 10MeV}\cdot{\rm fm}$ viscosity.}
	\label{fig:nonrel_energy_anim}
\end{center}
\end{figure}

\begin{figure}
\begin{center}
    	\includegraphics[width=1\textwidth]{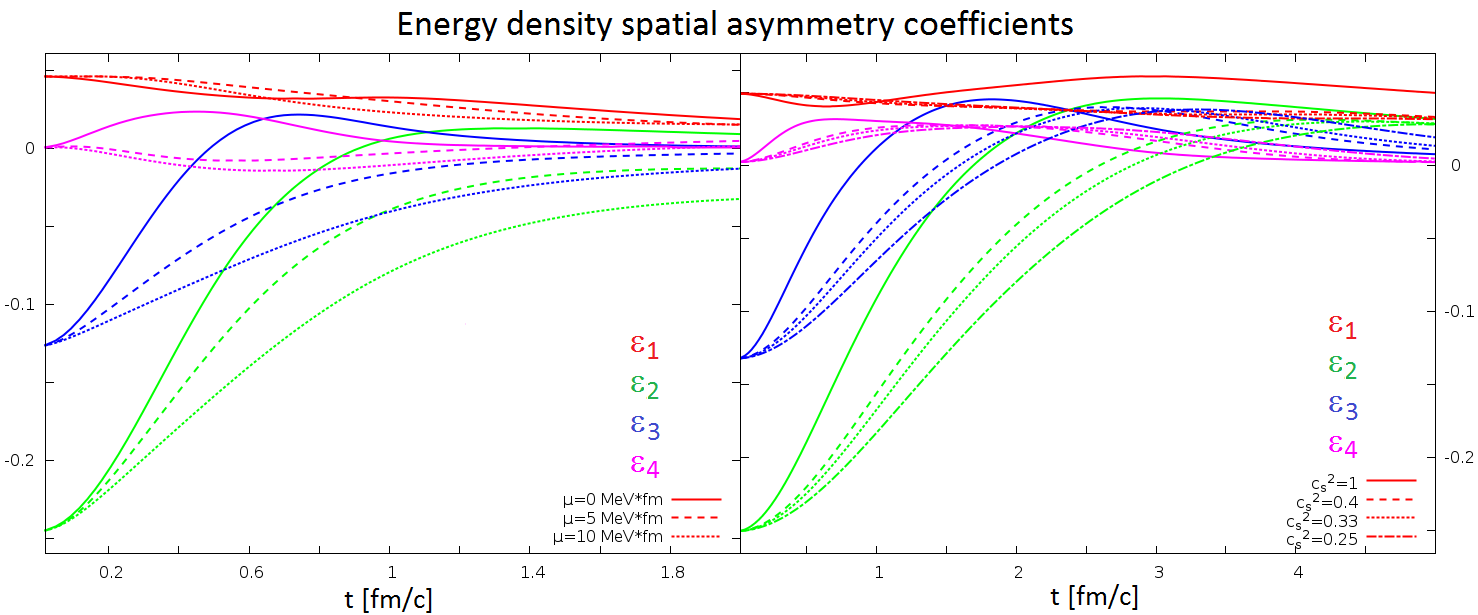}
	\caption{The time evolution of the asymmetry coefficients $\epsilon_{1,2,3,4}$ in the energy density,
	as modified by a small amount of
	viscosity (left plot) and the change in speed of sound (right plot).}
	\label{fig:nonrel_energy_cs2}
\end{center}
\end{figure}

\begin{figure}
\begin{center}
    	\includegraphics[width=0.7\textwidth]{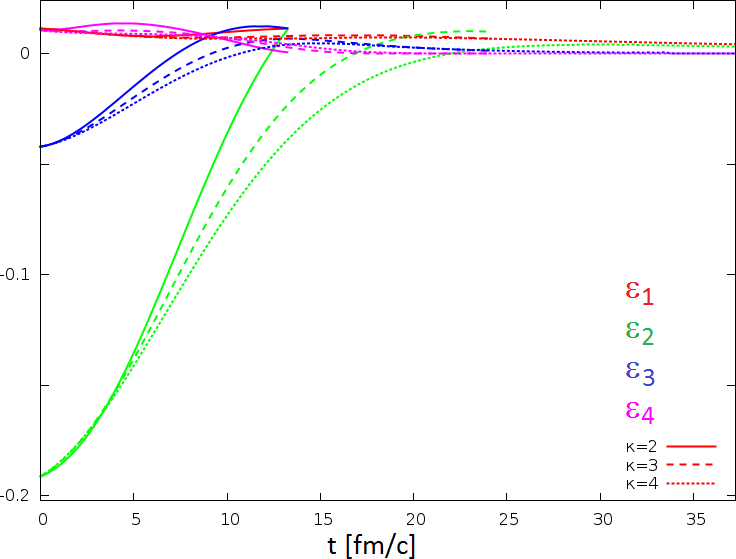}
	\caption{The time evolution of the asymmetry coefficients in the energy density in a relativistic calculation,
	as modified by the change in $\kappa=c_s^{-2}$ (right plot).}
	\label{fig:rel_energy_cs2}
\end{center}
\end{figure}

Fig.~\ref{fig:nonrel_energy_anim} shows the result of a nonrelativistic calculation of the time evolution of the energy density.
If we assume a small amount of viscosity, it makes the flow itself and thus the disappearance of asymmetries slower.
The time evolution of the asymmetries themselves is shown in Fig.~\ref{fig:nonrel_energy_cs2}. In this figure we also see
the effect of speed of sound in this nonrelativistic calculation: the reduction of speed of sound makes the asymmetries
disappear slower -- due to the reduction of the speed of sound waves. A similar effect is seen in case of a relativistic
calculation, as shown in Fig.~\ref{fig:rel_energy_cs2}: the increase of $\kappa=c_s^{-2}$ makes the disappearance of asymmetries
slower. It is important to see that this also slows down the speed of cooling, which means the system will freeze out latter. 

\section{Anisotropy mixing}
\begin{figure}[h!]
\centering
\includegraphics[width=1\textwidth]{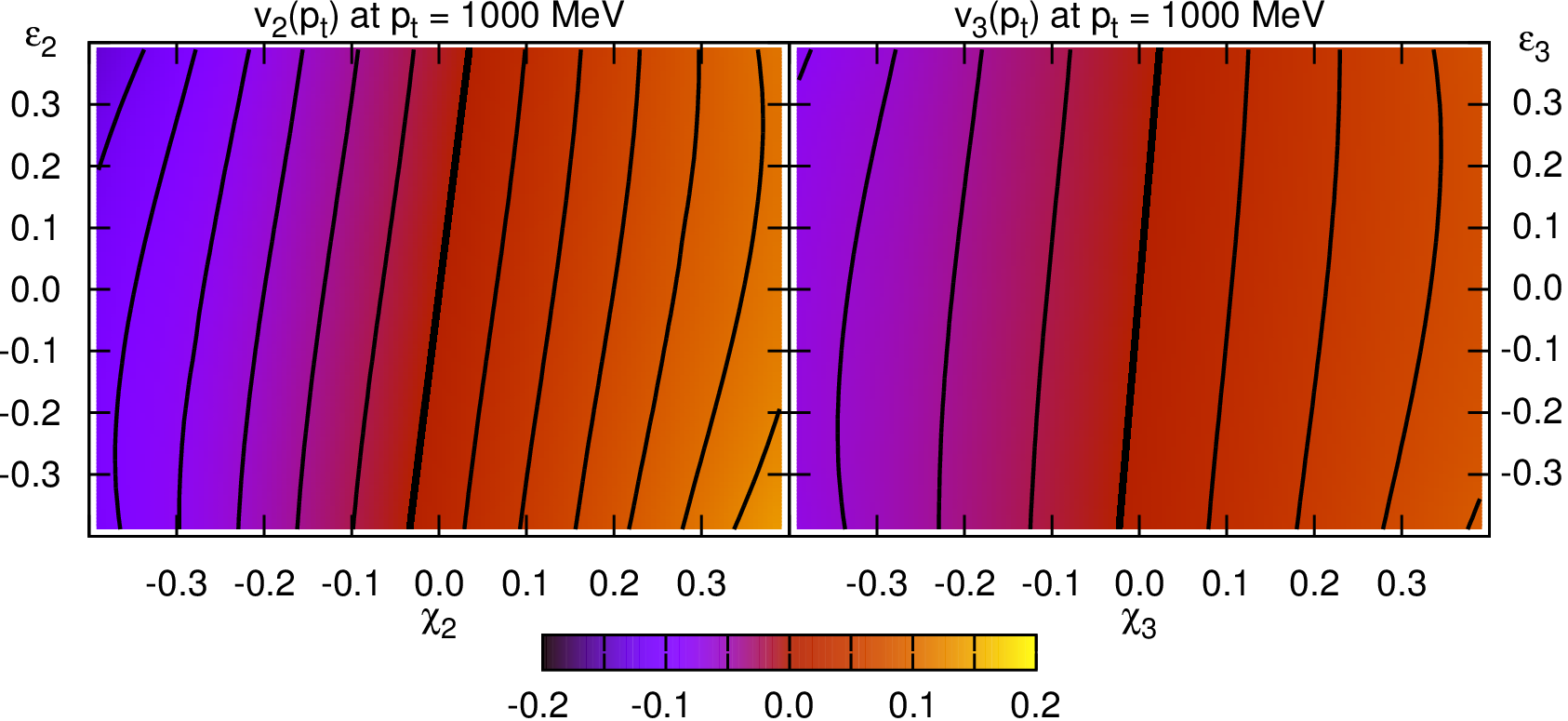}
\caption{\label{f:flows} Dependence of flow anisotropy coefficients $v_{2,3}$ on asymmetry
parameters $\chi_{2,3}$ and $\epsilon_{2,3}$. Velocity field asymmetry has a stronger
influence on flow coefficients.}
\end{figure}

It is important to see that there may be asymmetries in both momentum space (i.e. in the velocity field)
and in density (i.e. in energy density or pressure), and both influence the measured anisotropies. To investigate
this effect, we created a multipole version of the Buda-Lund model~\cite{Csanad:2003qa}, with a scale variable
given in Eq.~\ref{e:sNm}, but we introduced a multipole flow field as well. We start from a ``flow potential''
$\Phi$, which gives us the flow:
\begin{align}
{\bf v} = (\partial_x \Phi, \partial_y \Phi, \partial_z \Phi).
\end{align}
The flow field at a given time is spherically symmetric if $\Phi=\frac{r^2}{2H}$ with $H$ being a Hubble-coefficient
at that given time. Elliptical symmetry is obtained with $\Phi=\frac{r^2}{2H}(1+\chi_2\cos(2\varphi))$, while
\begin{align}
\Phi=\frac{r^2}{2H}(1+\chi_2\cos(2\varphi))+\frac{r^3}{3H^2}\chi_3\cos(3\varphi))
\end{align}
represents a triangular perturbation of the elliptical flow. Of course the various anisotropies can have various
event planes (symmetry planes), but the specific angle of these does not enter into the results.
Inspired by Ref.~\cite{Csanad:2008af}, we analyzed how $\chi_{2,3}$ and $\epsilon_{2,3}$ influence
flow coefficients $v_2$ and $v_3$ -- see results in Fig.~\ref{f:flows}. Compared to the spatial anisotropy, 
velocity field anisotropy has a much larger effect on elliptic and triangular flow coefficients.

\section{Summary and acknowledgments}
In this paper we showed an extension of the scope of analytic relativistic hydrodynamics to higher
order azimuthal asymmetries, compatible with realistic (event-by-event fluctuating) geometries.
Higher order flow observables were calculated from this model, and are found to be compatible with data. 
In the analytic model, anisotropy parameters were independent of time, thus we investigated their
time evolution in a numerical framework, developed for this purpose. We also investigated how 
velocity- and density-field anisotropies ``mix'', in the framework of a multipole Buda-Lund model.
We are thankful to Tamás Csörgõ and Márton Nagy for useful discussions with respect to this project.
We thank the WPCF community and the WPCF 2014 organizers, in particular the local hosts,
Tamás Novák and Tamás Csörgõ, for the possibility to present this work.
We also thankfully acknowledge the support of the OTKA grant NK 101438.

\bibliographystyle{../../../utphys}
\bibliography{../../../Master}

\end{document}